\documentclass{iaus}
\usepackage{graphicx}
\usepackage{epstopdf}

\title[Galactic PNe Abundance Patterns] 
{Galactic Abundance Patterns via Peimbert Types I \& II PNe}

\author[Milingo $\etal$]   
{J.B. Milingo$^1$, K.B. Kwitter$^2$, R.B.C. Henry$^3$, \and S.P. Souza$^2$}
\affiliation{$^1$Dept. of Physics \& Astronomy, Franklin \& Marshall College,
Lancaster, PA 17604, USA \break email: jmilingo@fandm.edu\\[\affilskip]
$^2$Dept. of Astronomy, Williams College, Williamstown, MA 01267 USA \break email: kkwitter@williams.edu, ssouza@williams.edu\\[\affilskip]
$^3$Dept. of Physics \& Astronomy, University of Oklahoma, Norman, OK 73019 USA \break email: henry@nhn.ou.edu}
\pubyear{2006}
\volume{234}  
\pagerange{119--126}
\date{?? and in revised form ??}
\jname{Planetary Nebulae in our Galaxy and Beyond}
\editors{M. J. Barlow \& R. H. M\'endez, eds.}
\begin{document}
\maketitle


\firstsection
\section{Introduction}

Planetary nebulae (PNe) are unique probes of chemical evolution. 
As shed envelopes resulting from the late evolutionary stages of
intermediate-mass stars $(1M_{\odot}{\le}M{\le}8M_{\odot})$, PNe are telling of both the
evolution of the progenitor and the natal ISM from which it formed.  On
a larger scale, the range in progenitor masses (hence main-sequence
lifetimes) and populous numbers of intermediate-mass stars make PNe useful signposts of how chemical composition varies spatially
and temporally across galaxies.  Observations of PNe
abundances also help constrain theoretical
predictions of how the initial chemical composition of intermediate-mass
stars is altered throughout the lifetime of the progenitor, mixed to the
surface and expelled, thus contributing to the chemical evolution of
their galactic environs.  For those elements that aren't altered during
the evolution of the progenitor, the natal ISM can be tested for the
yields of previous generations of massive and intermediate mass stars.

We present total element abundances based upon newly acquired spectrophotometry of a sample of $>$120 Galactic PNe (Table 1).  This new data set is extracted from spectra that extend from $\lambda$3600 - 9600\AA\ allowing the use of [S III] features at $\lambda\lambda$9069 \& 9532\AA.  Since a significant portion of S in PNe resides in S$^{+2}$ and higher ionization stages, including these strong features improves the extrapolation from observed ion abundances to total element abundance.  S is believed to be precluded from enhancement and depletion across the range of PNe progenitor masses making it an alternate metallicity tracer to the canonical oxygen.  If S can be reliably determined in PNe, its stability in intermediate mass stars makes it a valuable tool to probe the natal conditions as well as the evolution of PNe progenitors.  This is a continuation of our Type II PNe work, the impetus being to compile a relatively large set of line strengths and abundances with internally consistent observation, reduction, measurement, and abundance determination, minimizing systematic effects that come from compiling various data sets.  

With previous
observations of 85 Galactic PNe in hand we have recently added an
additional 40.  These PNe cover a substantial range in galactocentric distance, and
include Peimbert types I and II.  Peimbert ÒtypeÓ classifies PNe
according to chemical composition, a proxy for characteristics of the
progenitor star (Peimbert 1978).  This
compilation allows us to look for abundance patterns (total element ratios such as X/H and X/O) across PNe progenitor masses,
metallicities, and morphologies.  In addition to looking for signatures
of stellar evolution and nucleosynthesis via abundance patterns, we continue to explore the use of the near-IR
[S III] emission features as reliable indicators of S$^{+2}$ abundances,
improved extrapolated total sulfur, and its use as a metallicity tracer.


    %
\begin{table}\def~{\hphantom{0}}
  \begin{center}
  \caption{Mean values of O/H and X/O for our Galactic PNe and other samples.}
  \label{tab:XtoO}
  \begin{tabular}{llllll}\hline
      $O/H(x10^4)$  & $S/O(x10)$   &   $Cl/O(x10^3)$ & $Ar/O(x10^2)$ & $Ne/O$ & $Sample$ \\\hline
       4.41 $\pm$ 1.81   &  0.18 $\pm$ 0.09	& 0.43 $\pm$ 0.19 & 0.83 $\pm$ 0.46 & 0.28 $\pm$ 0.13 & Galactic Type I PNe$^a$\\	
       5.30 $\pm$ 1.97   &  0.12 $\pm$ 0.09 & 0.30 $\pm$ 0.14 & 0.52 $\pm$ 0.25 & 0.23 $\pm$ 0.08 & Galactic Type II PNe$^a$\\
       5.05 $\pm$ 1.96   &  0.13 $\pm$ 0.09	& 0.34 $\pm$ 0.17 & 0.60 $\pm$ 0.35 & 0.25 $\pm$ 0.10	& all PNe in our sample$^a$\\
       3.05 $\pm$ 2.63   &  0.26 $\pm$ 0.21	& $\cdots$ & 0.64 $\pm$ 0.29 & 0.21 $\pm$ 0.14 & Galactic PNe$^b$\\
       4.8~ $\pm$ 2.0   &  0.17 $\pm$ 0.14 & $\cdots$ & 0.48 $\pm$ 0.48 & $\cdots$ & Galactic PNe$^c$\\
       4.4~ $\pm$ 0.19 &  0.25 $\pm$ 0.02 & 0.47 $\pm$ 0.04 & 0.69 $\pm$ 0.05	 & $\cdots$ & Galactic PNe$^d$\\
       4.57 & 0.32 & 0.69 & 0.33 & 0.15 & Solar values$^e$ \\
       5.25 &  0.28 & 0.41 & 0.59 & $\cdots$ & Orion (gas + dust)$^f$ \\
       2.11 $\pm$ 1.11 &  0.29 $\pm$ 0.04 & $\cdots$ & 0.62 $\pm$ 0.11 & 0.21 $\pm$ 0.03 & M101 HII regions$^g$\\
       $\cdots$ & 0.36 & 0.50 & 0.89 & $\cdots$ & MW H II regions$^h$ \\\hline
  \end{tabular}
 \end{center}
 $^a$\cite[Kwitter \etal\ (2001)]{Paper I}, \cite[Milingo \etal\ (2002)]{Paper IIb}, \cite[Henry \etal\ (2004)]{Paper IV}, 
 $^b$\cite[Maciel \& K\"{o}ppen (1994)]{Maciel94}, 
 $^c$\cite[Kingsburgh \& Barlow (1994)]{Kingsburgh94}, 
 $^d$\cite[Aller \& Keyes (1987)]{Aller87}, 
 $^e$\cite[Asplund \etal\ (2005)]{Asplund05}, 
 $^f$\cite[Esteban \etal\ (1998)]{Esteban98}, 
 $^g$\cite[Kennicutt \etal\ (2003)]{Kennicutt03}, 
 $^h$\cite[Rodriguez (1999)]{Rodriguez99}\\
\end{table}
\firstsection
\section{Ongoing Work}\label{sec:concl}
We now have abundances for
$>$120 Galactic PNe.  This data is unique in that it is based upon
newly acquired spectrophotometry covering an extended range in wavelength.  Utilizing a 5-level atom
abundance routine we've carefully determined ${T_e}, {N_e}$, and ICFs, providing a
consistent and homogeneous set of data.  We are looking to minimize
systematic effects that may creep in when combining various samples that
utilize different reduction and abundance determination schemes, thus disguising subtle
abundance patterns such as enhancements or depletions due to
nucleosynthesis.  Further analysis needs to be done
to discern scatter due to uncertainty from true abundance distinctions and breadth.  For example the anomalously low S/O ratio for PNe begs further examination (see R.B.C. Henry $\etal$ in these proceedings), trends in N/O could signal the ON cycle at work, and the breadth in Ne/O for Type I PNe, due to a few extreme outliers, could be illustrating neon enrichment.
In looking for distinguishing characteristics within our
abundance data, more Type I PNe need to be added, and the entire ensemble of data requires a rigorous statistical analysis.

We gratefully acknowledge support from the AAS Small Research Grants program, the Franklin \& Marshall Committee on Grants, and NSF grant AST-0307118.


\begin{thebibliography}{}

\bibitem[Aller \& Keyes (1987)]{Aller87}
     {Aller, L.H. \& Keyes, C.D.} 1987,
     \textit{ApJS} 65, 405

\bibitem[Asplund \etal\ (2005)]{Asplund05}
     {Asplund, M., Grevesse, N., Sauval, A.J.} 2005,
     in: T.G. Barnes III \& F.N. Bash (eds.),
     \textit{Cosmic Abundances as Records of Stellar Evolution and Nucleosynthesis}, 
     ASP Conference Series Vol. 336 (San Francisco: ASP), p.\ 25

\bibitem[Esteban \etal\ (1998)]{Esteban98}
     {Esteban, C., Peimebert, M., Torres-Peimbert, S., \& Escalante, V.} 1998,
     \textit{MNRAS} 295, 401

\bibitem[Henry \etal\ (2004)]{Paper IV}
     {Henry, R.B.C.,Kwitter, K.B., Balick, B.} 2004,
     \textit{AJ} 127, 2284

\bibitem[Kennicutt \etal\ (2003)]{Kennicutt03}
     {Kennicutt, R.C., Jr., Bresolin, F., \& Garnett, D.R.} 2003,  
     \textit{ApJ} 591, 801

\bibitem[Kingsburgh \& Barlow (1994)]{Kingsburgh94}
     {Kingsburgh, R.L. \& Barlow, M.J.} 1994,
     \textit{MNRAS} 271, 257

 \bibitem[Kwitter \etal\ (2001)]{Paper I}
     {Kwitter, K.B., \& Henry, R.B.C.} 2001, 
     \textit{ApJ} 562, 804

  \bibitem[Kwitter \etal\ (2003)]{Paper III}
     {Kwitter, K.B., Henry, R.B.C., \& Milingo, J.B.} 2003,  
     \textit{PASP} 115, 80

\bibitem[Maciel \& K\"{o}ppen (1994)]{Maciel94}
     {Maciel, W.J. \& K\"{o}ppen, J.} 1994,
     \textit{A\&A} 282, 436

  \bibitem[Milingo \etal\ (2002)]{Paper IIa}
     {Milingo, J.B., Kwitter, K.B., Henry, R.B.C., \& Cohen, R.E.} 2002,  
     \textit{ApJS} 138, 279

  \bibitem[Milingo \etal\ (2002)]{Paper IIb}
     {Milingo, J.B., Henry, R.B.C., \& Kwitter, K.B.} 2002,  
     \textit{ApJS} 138, 285

  \bibitem[Peimbert (1978)]{Peimbert78}
     {Peimbert, M.} 1978,  
     in: Y. Terzian (ed.),
     \textit{Planetary Nebulae: Observations and Theory}, 
     Proc. IAU Symposium No. 76 (Dordrecht: Reidel), p.\ 215

\bibitem[Rodriguez (1999)]{Rodriguez99}
     {Rodriguez, M.} 1999
      \textit{A\&A} 351, 1075 

\end{thebibliography}
\end{document}